# Omori's law in the Internet traffic

SUMIYOSHI ABE[1] and NORIKAZU SUZUKI[2]

[1]*Institute of Physics, University of Tsukuba, Ibaraki 305-8571, Japan*
[2]*College of Science and Technology, Nihon University,*
 *Funabashi, Chiba 274-8501, Japan*

**Abstract**. — The Internet is a complex system, whose temporal behavior is highly nonstationary and exhibits sudden drastic changes regarded as main shocks or catastrophes. Here, analyzing a set of time series data of round-trip time measured in echo experiment with the Ping Command, the property of "aftershocks" (i.e., catastrophes of smaller scales) after a main shock is studied. It is found that the aftershocks obey Omori's law . Thus, the Internet shares with earthquakes and financial market crashes a common scale-invariant feature in the temporal patterns of aftershocks.

The Internet is a typical complex random network, which has a self-organizing hierarchical structure, highly intricate tangle and connection diversity [1]. It exhibits several kinds of scale invariant behaviors, which may be essential for the network to be resilient and robust to random errors, breakdowns and attacks [2-4]. For instance, a power-law distribution of connectivity has recently been derived analytically in Ref. [5] by using random graph theory. In addition, it has been observed [6] that a time series describing the states of congestion can be well characterized by the scale-invariant $q$-exponential distributions, i.e., generalizations of the Zipf-Mandelbrot power-law distribution [7], which maximize the Tsallis entropy [8] under appropriate constraints on the averages of observable quantities. (In this respect, it may also be worth mentioning that the Albert-Barabási analytic solution in Ref. [5] is actually the $q$-exponential distribution.)

In the nonstationary Internet time series, there are two separate time scales: one is the user's long time scale (typically ~ 1 hour) and the other is a short time scale (~ 100 ms) associated with drastic changes of the state, which may be identified with shocks or catastrophes. This feature seems to be in common with phenomena of earthquakes, in which external driving or injection of energy may last for many years and the internal relaxation needs a few seconds or minutes. Such a similarity naturally leads to a question if relatively smaller aftershocks temporally distribute according to Omori's law [9]. In this context, it may be worth mentioning that Omori's law is also found in financial time series [10].

In this Letter, we present an experimental evidence for validity of Omori's law in the nonstationary Internet time series obtained by performing echo experiment. This result indicates that the Internet contains long-time memory of past events which influences the states in the future, and the slow power-law decay represents scale invariance of another type in the Internet traffic.

Echo experiment we have performed uses the Ping Command [11,12]. A Ping signal is emitted from a local host computer, takes a round trip to a destination host and returns to the original host through ten odd routers, through which the signal paths are connected to the whole network in the complex manner. The signals are emitted one after another with the fixed time interval of 1s. The whole collected data of the round-trip times defines a time series. The Internet time series thus obtained is highly nonstationary and contains sudden drastic changes in its temporal pattern with a short time scale, which may be regarded as main shocks.

Omori's law is an empirical one found in the temporal pattern (i.e., waiting time) of aftershocks associated with a significant earthquake. This law states that the number of aftershocks, $dN(t)$, occurring in the time interval, $(t, t+dt)$, after a main shock at $t=0$ obeys

$$\frac{dN(t)}{dt} \sim \frac{1}{t^p} \qquad (p>0). \qquad (1)$$

The observed data of earthquakes show that the value of the exponent, $p$, ranges from 0.9 to 1.5.

As we shall see in the Internet time series, so far only the cases of $p < 1$ are observed. In such cases, the cumulative number of aftershocks occurred until time $t$ after a main shock is written as

$$N(t) = A t^{1-p}, \qquad (2)$$

where $A$ is a constant.

In Fig. 1, from extensive tests, we present a typical example of the observed Internet time series of round-trip times of the Ping signals. A main shock may be recognized at $t = 0$. A nonstationary regime, on which we focus our attention, is indicated by the horizontal arrow. In Fig. 2a, the log-log plots of the elapsed time, $t$, after the main shock and the cumulative number of aftershocks are shown. The straight lines are drawn by making use of the method of least squares. Three different values of the threshold, which make it possible to identify aftershocks, are examined. The values of the correlation coefficient are large (~ 0.97). In Fig. 2b, the cumulative number of aftershocks is plotted with respect to time, $t$, on the linear scale. The larger the threshold value is, the better fitting by Omori's law is. In general, we observe the oscillatory behavior around Omori's law.

In conclusion, we have presented an experimental evidence for relevance of Omori's law to the Internet traffic. The observed values of the exponent, $p$, are roughly $p < 0.8$ and the case of $p > 1$ has not been recognized throughout our extensive tests so far. The present result means that the Internet shares with earthquakes and financial market crashes a common scale-invariant feature in the temporal patterns of aftershocks. Thus, Omori's law may be expected to be universal for complex systems and their nonstationary temporal behaviors.

The oscillatory behavior around Omori's law is an intriguing phenomenon in analogy with earthquakes. In Ref. [13], the log-periodic nature of oscillation around Omori's law for earthquake aftershocks has been noticed, and therefore it would be of interest to further clarify the property of the oscillatory behavior in the Internet traffic. Also in analogy with earthquakes, it should be examined if another scaling law of the Gutenberg-Richter-type holds for the Internet shocks. These points will be discussed elsewhere [14].

# Figure Captions

Fig. 1  The observed Internet time series of round-trip times of the Ping signals. The main shock is located at $t = 0$. The local host computer and the destination host are ns2.phys.ge.cst.nihon-u.ac.jp (133.43.112.51) and www.th.phys.titech.ac.jp (131.112.122.76), respectively. The number of intermediate routers is 16. The experiment was performed from 9:50:35 to 12:07:49 on 13 May, 2002. The number of the data points in the regime to be analyzed, which is indicated by the horizontal arrow, is 8223. The packet dropping rate is 0.15%.

Fig. 2  **a**. The log-log plots of the elapsed time, $t$, after the main shock at $t = 0$ and the cumulative number of aftershocks above the threshold value defined here as $m + \alpha\sigma$ with $m = 147.72$ and $\sigma = 63.60$ respectively being the mean value and the standard deviation of round-trip time in the regime under consideration. Three values of $\alpha$ are examined: (1) $\alpha = 3.5$, (2) $\alpha = 3.8$ and (3) $\alpha = 3.9$. The straight lines representing Omori's law are drawn by making use of the method of least squares. The values of $p$ and $A$ are: (1) $p = 0.68$, $A = 1.6$, (2) $p = 0.72$, $A = 1.1$ and (3) $p = 0.77$, $A = 0.94$. The values of the correlation coefficient, $\rho$, are: (1) $\rho = 0.968$, (2) $\rho = 0.996$ and

(3) $\rho = 0.987$.

**b**. The plots of the cumulative number $N(t)$ with respect to $t$ on the linear scale, corresponding to the log-log plots in **a**.

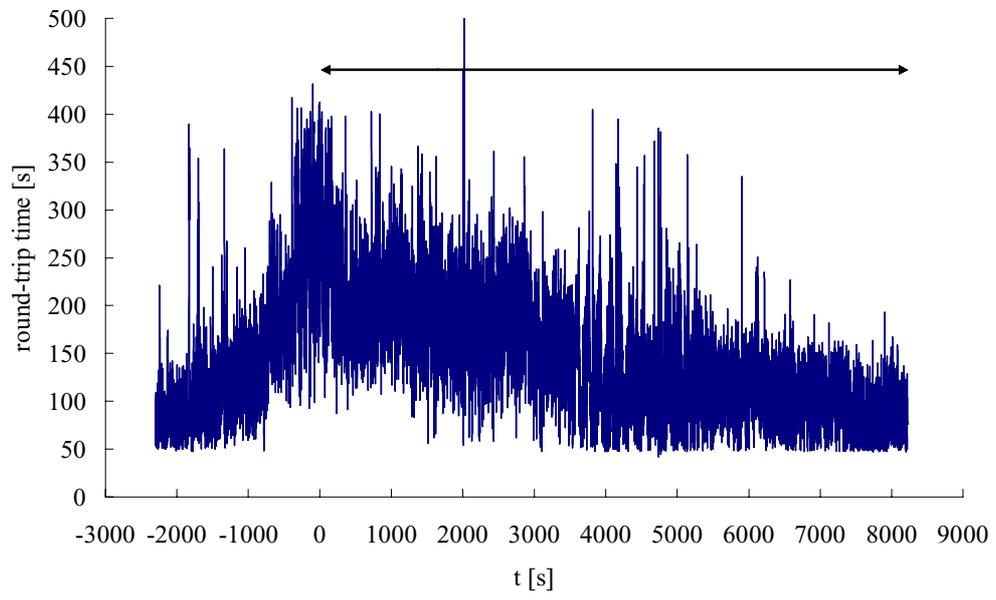

Fig. 1

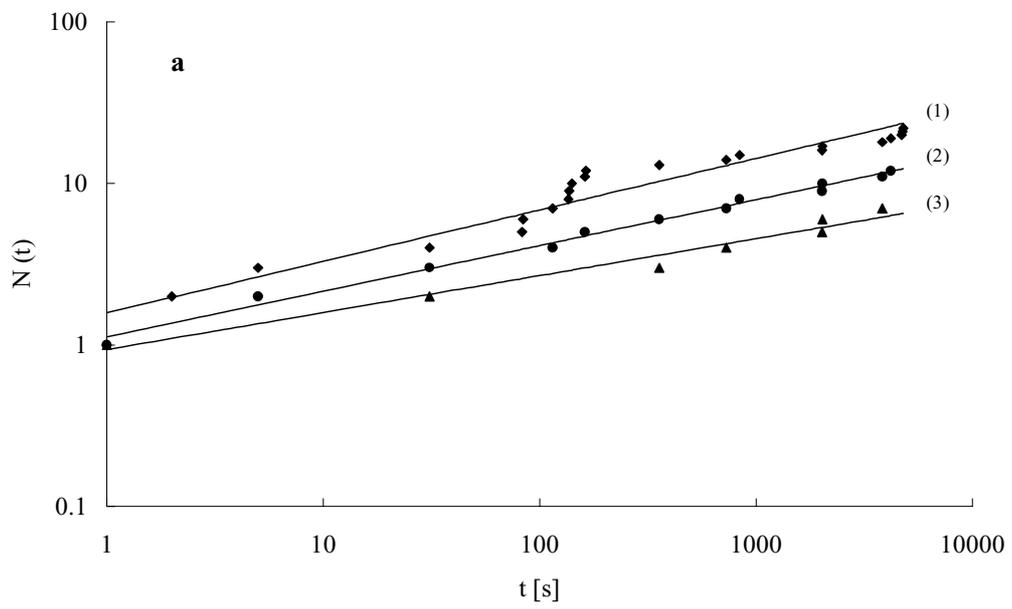

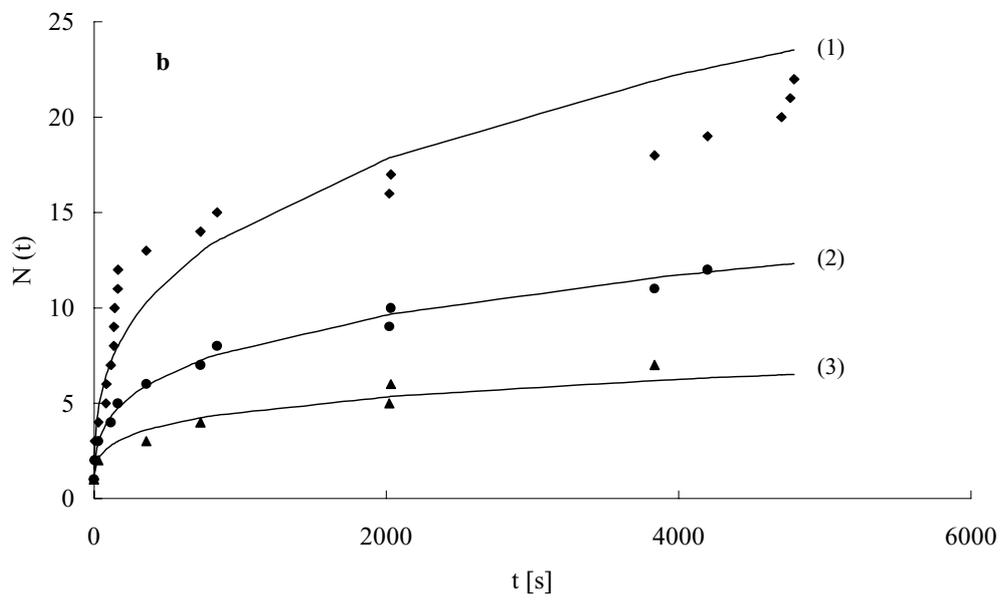

Fig. 2